\documentstyle[aps,prl,psfig,epsfig,amstex]{revtex}

\begin{document}

\title{Anomalous quantum Hall conductivity and resonances 
in coupled layers.}

\author{A. L. Chudnosvkiy and S. E. Ulloa} 

\address{Department of Physics and Astronomy and Condensed Matter and
Surface Sciences Program \\ Ohio University, Athens, Ohio 45701--2979}

\maketitle

\begin{abstract}
A two-layer system coupled via tunneling and with different carrier
masses in each layer is investigated in the integer quantum Hall
regime.  Striking deviations of the one-layer Hall conductivity from
the usual quantization are found, if resonance between Landau levels
of different layers is achieved. The appearance of {\it negative}
jumps in the Hall conductivity is also predicted under suitable
conditions.  The results obtained are robust against disorder in the
system.
\end{abstract}


Recent progress in nanotechnology has made possible the creation of
electronic systems with unusual and interesting properties.  Prominent
examples include spatially separated electronic  layers coupled by
tunneling \cite{Eisenstein} or only via Coulomb interactions 
\cite{Gramila}, quantum dot arrays in different configurations
\cite{Alivisatos,Petroff}, or even combinations of layers and quantum
dots \cite{Main,Ensslin}.  These artificial systems provide 
interesting analogues to natural systems, 
although with different
characteristic length and energy scales.  On the other hand,  the artificially
constructed systems may serve as a laboratory to discover and utilize in practice 
new physical
effects, hardly seen in naturally occurring solids.

The aim of this paper is to draw attention to qualitatively new 
physical effects which are predicted to occur in the integer quantum 
Hall (QH) regime for a system of two conducting layers coupled via 
tunneling and with charge carriers of different effective masses. 
In the inset to Fig.\ \ref{figexp}, we suggest a suitable experimental 
configuration for such a system. 
The spatial separation of the layers allows for experimental measurements 
of the quantum Hall resistance of each layer {\it independently}, 
which would exhibit  
the well-known plateau structure versus the magnetic field 
(or the Fermi energy).  The presence 
of the other layer coupled via tunneling, however, leads to quantum 
interference 
effects that result in the deviation of the resistivity  
plateaus from quantized values, if two Landau 
levels of the different layers happen to be close in energy (in resonance). 
  
By applying a bias voltage $V_0$, as shown in Fig.\ \ref{figexp}, 
one can "slide" the system of LL's of one layer with 
respect to that of the other, thus changing the value of the one-layer resistivity 
on the plateau, and producing a tunable quantum Hall resistor.
One should mention that 
such `gated' configurations have been successfully implemented in a 
number of clever magnetocapacitance and transport experiments
\cite{Eisenstein,Ensslin,Miller}, and are in great part the motivation 
for this work.
Different realizations can also be proposed and are
experimentally feasible, including 
one with dopants of different kind (donors
and acceptors) on each layer, as electrons and holes would have
substantially different masses (similar systems have been recently
studied \cite{Sivan-ICPS}), or one with different semiconductor
materials across the interface, such as InAs-GaAs, to give
different-mass carriers in the conduction bands.  One of the
particularly interesting realizations of such two-layer system
 is a two-dimensional electron system (2DES),
coupled via a tunneling barrier to a planar array of quantum dots
(QDs) \cite{Ensslin}.  
In the limit of 
close in-plane separation between dots, to allow direct electron 
tunneling, the 2DES-QDs structure would effectively reduce to 
that of two layers with different effective masses.  

In a two-layer system, the longitudinal and transverse resistivities 
become matrices  with components reflecting the position of the measuring
contacts, for example, $\rho_{xy}^{\alpha\beta,\gamma\delta}$, where
$\alpha,\beta,\gamma,\delta$ can be either $l$ or $h$, denoting the 
`light-' or `heavy-mass' 
particle  layer, as shown in the inset  to Fig.\ \ref{figres1}(a). 
It is important, that each component of the resistivity tensor can be measured 
independently by switching the measuring contacts. In what follows, we shall 
discuss the properties of the conductivity matrix 
$\sigma_{xy}^{\alpha\beta,\gamma\delta}=(\rho^{-1}_{xy})^{\alpha\beta,\gamma\delta}$, 
which can be calculated directly from the experimental data. 
Similarly, one can obtain any desired $\rho_{xy}$ component from the 
theoretical calculations of $\sigma_{xy}$, as we illustrate below. 

Under the resonant conditions above, the 
components of the 
transverse conductivity matrix  exhibit striking behavior, 
as the Fermi level of the system moves through the level structure.  
First, the jumps deviate from the usual value, producing  
Hall conductivity plateaus at {\it non-quantized values}.  
Moreover, the jumps can be both positive and {\it negative} --- 
that is, the value of the transverse 
conductivity measured in a given layer can change non-monotonically 
(see Fig.\ \ref{figres1}(b)).  
Finally, if the layers have different 
effective mobilities (with corresponding different widths of their
LL's), singularities in the density of
states may occur, leading in turn to additional features in the 
components of the Hall conductivity in each layer. 
The prediction of non-quantized and even negative jumps of the
transverse conductivity in the two-layer QH system under resonant
conditions is one of the main results of this paper.  

One can understand the deviation from exact quantization in each 
layer as resulting from the shift of LL energies due to the 
coupling between the layers. In a clean system, the
position of the states is shifted due to the interlayer tunneling 
$t$, as defined by the well-known expression,
 \begin{equation}
\epsilon_k^{\pm}=\frac{1}{2}\left(E_k^l+E_k^h\pm
\left[(E_k^l-E_k^h)^2+4t^2\right]^{1/2}\right),
 \label{shiftLL}
 \end{equation}
where $E_k^l=(k+1/2)\Omega$ is the energy of the unperturbed (no tunneling) 
LL in the 
layer of light particles, while $\Omega$ denotes the cyclotron 
frequency in this "l-layer", 
$E_k^h=E_0+(k+1/2)\Omega/M$  is the energy of the 
unperturbed LL in the heavy particle layer ($M$ is the mass ratio between 
heavy and light particles, and $E_0$ is the energy offset between the two
systems of LL's due to the voltage applied). Here and in what follows 
we use units with $h=1$ and $m_e=1$.
Using the Kubo formula, up to corrections of order $t/\Omega$,
one obtains the following expression for the jump in transverse 
conductivity around the n-th Landau level in the l-layer 
\begin{equation}
\Delta_n\sigma_{xy}^{ll,ll}=\lim_{\omega\rightarrow 0}
\frac{\Omega^2}{2\omega}\left[\frac{1}{\epsilon^l_{n+1}-
\epsilon^l_{n}-\omega}
- \frac{1}{\epsilon^l_{n+1}-\epsilon^l_{n}+\omega}\right].
\label{jumpsimple}
\end{equation}In the usual case, this expression yields one conductivity  quantum, 
if the degeneracy of the Landau level (which equals $\Omega$ in 
these units) is equal to the separation between neighbor LL's, 
$\epsilon^l_{n+1}-\epsilon^l_{n}$, as is the case in the absence of  
tunneling. The tunneling between the layers, however, leads 
to  $\epsilon^l_{n+1}- \epsilon^l_{n}\neq\Omega$, 
and hence to the deviation of the jump from one conductivity 
quantum of order $t/\Omega$. 
Another effect of  tunneling is perhaps more subtle, as the
conductivity involves in general not only
 LL's of the same layer, but also those of
 {\it different} layers.  This is reflected in the presence 
of energy differences of the type $\epsilon^l_{n+1}- 
\epsilon^h_n$ in the general 
expression for the Hall conductivity (not shown). 
Sliding one system of
LL's with respect to the other, one can achieve the resonance 
between the energies $\epsilon^l_{n+1}$ and $\epsilon^h_n$. 
Close to
that resonance, additional {\it nonquantized} jumps in the 
transverse conductivity of each layer appear, the sign of the 
jumps being defined by the sign of the difference 
$\epsilon^l_{n+1}- \epsilon^h_n$. 
The jumps can, therefore, be {\it negative}, as well as positive. 

One should emphasize, however, that the topo\-logically-\-expected
quantization condition is fulfilled for the combination 
\begin{equation}
\sigma_{xy}^{ll,ll}+\sigma_{xy}^{ll,hh}+\sigma_{xy}^{hh,hh}+
\sigma_{xy}^{hh,ll}.
\label{quant}
\end{equation} 
This quantization  
conditions follows naturally, if one considers a two-layer system, 
connected fully in parallel. The resulting 
conductivity  of the effective one layer system should be quantized 
in the traditional way, and it 
remains unaffected by the singularities due to disorder.

At large tunneling ($t\sim \Omega$), the components $\sigma_{xy}^{ll,ll}$ 
and $\sigma_{xy}^{hh,hh}$ cease to reflect the measured conductances in 
each layer because of the strong interlayer coupling. In the 
limit of very large tunneling, the measurements would give only the 
effective layer results, given by the trace of the Hall conductivity tensor  
and always well-quantized.  

To obtain a full analytical description of the system considered, we 
generalize the replica nonlinear sigma model approach to the 
integer QH effect developed by Pruisken \cite{Pruisken} to the 
case of a two-layer system with different effective masses. 
The Hamiltonian of the two-layer model we are interested in 
can be written as 
\begin{eqnarray}
\nonumber 
H=&&\sum_r \left\{\hat{a}_r^+
(\hat{\pi}_\mu\hat{\pi}^\mu+V_l) \hat{a}_r + \hat{c}_r^+
[E_0+\frac{1}{M}\hat{\pi}_\mu\hat{\pi}^\mu+V_h]\hat{c}_r \right. \\ 
&& \left. + t\left(\hat{a}_r^+\hat{c}_r +
\hat{c}_r^+\hat{a}_r\right)\right\},
\label{Ham}
\end{eqnarray}
 where the fermionic operators $\hat{a}$ and $\hat{c}$ refer to the
states in the two different layers, the operator $\hat{\pi}_\mu
\hat{\pi^\mu}$ describes the kinetic energy in a magnetic field
($\hat{\pi}^\mu\equiv -i\nabla_\mu-A_\mu$, $\mu=x,y$,
$\hat{\pi}_\mu=(\hat{\pi}^\mu)^+)$, $t$ is the tunneling matrix element
between the two layers, and $E_0$ describes the relative position of 
the bottom of the band of layer $c$ with respect to the layer $a$.  The
mass of the electrons in layer $a$ is taken as unity, so that $M \geq
1$ is the relative mass of the `heavier' quasiparticles in layer $c$.  
The random gaussian-distributed potentials $V_l(r)$ and $V_h(r)$ 
describe
the disordered scattering in the $l$-, and the $h$-layer, 
respectively. The
only nonvanishing correlators of the random potentials are 
$
 \langle V_\alpha(r)V_\alpha(r') \rangle
 =\frac{1}{2\pi\nu\tau_\alpha}\delta(r-r')
$, 
where $\nu$ is the density of states of a LL, 
$\alpha = l, h$, and $\tau_l$ and $\tau_h$ are the mean free 
times for a particle
in the light- and heavy-mass layers, respectively 
($1/\tau$ is the LL width).
Note, that we consider here a model without possible disorder in the
tunneling between the layers, $t$. The model with disorder in the 
tunneling belongs to a different class and will be considered in the 
future. 

Following the ideas of Ref. \cite{Pruisken}, we introduce in 
the two-layer system the basis of grassmanian vectors 
$\Phi=(\Psi_l, \Psi_h)^T$,
where the subscripts $l$ and $h$ relate to the layers.  Each of the
grassmanian components consists in turn of two variables, relating to
the advanced and retarded sectors, $\Psi_{l,h}=(\psi_{l,h}^A,
\psi_{l,h}^R)$.

Employing the replica trick to average over the disorder, we decouple
the resulting four-fermionic terms with the help of the
Hubbard-Stratonovich transformation, and integrate out the fermionic
degrees of freedom. 
The decoupling matrix field has the structure
 \begin{equation}
  {\bf \hat{Q}}=\left(
  \begin{array}{cc} 
\hat{Q} & 0 \\ 0 & \hat{R} 
\end{array}\right),
 \end{equation}
where the $2n\times 2n$ hermitian matrices $\hat{Q}$ and $\hat{R}$
facilitate the decoupling  of the four-fermionic terms in the `light' 
and `heavy' sectors, respectively.
According to the standard procedure \cite{Pruisken}, we look for a
mean-field solution for the matrix fields $\hat{Q}$ and $\hat{R}$ of
the form $\hat{Q}=q{\bf\sigma^z}\otimes {\bf 1_n}$, 
$\hat{R}=r{\bf\sigma^z}\otimes {\bf 1_n}$, where $\bf\sigma^z$ 
is a Pauli matrix. 
The mean-field
equations for $q$ and $r$ are then obtained in the form 
 \begin{eqnarray} 
 \nonumber && q =\sum_k\frac{1}{D_k}\left\{\frac{r}{2\tau_h}
\left[\frac{q r}{4\tau_l\tau_h}-(E^l_k-\epsilon_F)(E^h_k-\epsilon_F)+t^2
\right] \right.\\ && \left.
+(E^h_k-\epsilon_F)\left[\frac{q}{2\tau_l}(E^h_k-\epsilon_F)
+\frac{r}{2\tau_h}(E^l_k-\epsilon_F)\right]\right\}, \label{q}
 \end{eqnarray}
 where
 \begin{eqnarray}
\nonumber
&& D_k=\left[\frac{q r}{4\tau_l\tau_h}-
 (E^l_k-\epsilon_F)(E^h_k-\epsilon_F)+
t^2\right]^2 \\
&& +\left[\frac{q}{2\tau_l}(E^h_k-\epsilon_F)
+\frac{r}{2\tau_h}(E^l_k-\epsilon_F)\right]^2,
 \end{eqnarray}
and with a similar equation for $r$.
In the sum over the $k$-Landau levels, only the contribution of
the level closest to the Fermi energy, $\epsilon_F$, should be kept. 

Solving Eq.\ (\ref{q}) and the analogous one for $r$ numerically, 
we can calculate
the transverse conductivity tensor $\hat{\sigma}_{xy}$.  Here, we
concentrate on the behavior of the diagonal components of the
transverse conductivity tensor, which correspond to the measurements of
the conductivity when transverse 
contacts are applied in each layer, while the current is injected
into both layers (see inset in Fig.\ \ref{figres1}(a)).  If there is
no resonance between LL's of different layers, the behavior of the
transverse conductivity in each layer is quite normal, as the
conductivity jumps by one quantum ($e^2/h$) each time the Fermi energy
passes a Landau level of the given layer.

On the other hand, if there is a resonance between the LL's of
different layers, the behavior of the transverse conductivity changes
drastically, exhibiting different regimes as function of the disorder
strength in different layers.  There are two types of resonances with
different behavior of $\hat{\sigma}_{xy}$: {\bf (i)} resonance between LL's
with the same quantum number, $E_n^l=E_n^h$; and {\bf (ii)} resonance
between LL's with quantum numbers differing by 1, $E_n^l=E_{n\pm 1}^h$.

Both types of resonance are affected by the disorder, and in a
singular way for strong disorder. 
Weak disorder will produce level broadening and a slight shift of
the eigenenergies of the clean system, given by Eq. (\ref{shiftLL}), 
but their splitting is still proportional to $t$ on
resonance, and the corresponding states are mixtures of the light- 
and heavy-particle layer eigenfunctions.

In case {\bf (i)} and weak disorder, as shown in Fig.\ \ref{figres1}(a), 
the transverse conductivity components, $\sigma_{xy}^{ll,ll}$, 
$\sigma_{xy}^{hh,ll}=\sigma_{xy}^{ll,hh}$ and
$\sigma_{xy}^{hh,hh}$, experience a jump each time the Fermi level
crosses one of the two eigenenergies of the spectrum, $\epsilon_n^l$
and $\epsilon_n^h$.  Notice that due to the tunneling coupling between
the layers, these eigenvalues are shifted with respect to the resonant
energy $E_n^l=E_n^h$ by an amount proportional to $t$.  The jumps of
the transverse conductivity measured in each layer are {\it
non-quantized}, with plateaus in each layer not
equal to an integer number of quanta.  Notice, however, that
the {\it sum} of the values of the Hall conductivity in the two layers,
corresponding to combination (\ref{quant}) is 
indeed {\it well-quantized}, as one would expect of the entire system 
based on general topological considerations \cite{Laughlin,Kohmoto}.  
The departure from quantization of the conductivity plateaus in different 
layers grows with the difference in effective mass of the
particles, with the higher conductivity plateau for the lighter
particles.  The width of the non-quantized plateau in Fig.\
\ref{figres1}(a) grows with the tunneling amplitude $t$ causing the
splitting of the resonant LL's. The nonquantized plateau of the components 
of the conductivity imply a similarly nonquantized plateaus of the 
components of the resistivity, which would be explicitly observable in 
experiment (as an example, the dependence of $\rho _{xy}^{ll,ll}$ is 
shown in Figure \ref{figres1}(a)).

In the case {\bf (ii)} of a resonance with different Landau index, the
deviation of the transverse conductivity from the usual behavior is
even more profound.  Figure \ref{figres1}(b) shows the dependence of 
the transverse conductivity in each layer vs.\ Fermi energy position, in
the case of the resonance $E_n^l=E_{n+1}^h+\delta$. A small deviation 
from the exact resonance is needed in order to see the middle plateau 
in Fig. \ref{figres1}(b) between the two nonquantized jumps, 
otherwise (for $\delta=0$) 
one would see only one jump (also nonquantized), which is the sum of the 
jumps around the levels $E_{n+1}^h$ and $E_n^l$.  Here, similar to the
previous case, the transverse conductivity in each layer jumps by {\it
non-quantized} values when the Fermi energy crosses one of the shifted
energy levels, $\epsilon_n^l$ and $\epsilon_{n+1}^h$.  The jump of each
component can, however, exceed one quantum, in which case the jump in
the other layer will be {\it negative}, as to ensure the proper
quantization of the conductivity trace.  The amplitude of the jump of
each component grows with the tunneling strength $t$, as shown in the
inset. 
The tunneling determines also the width of the resonant region.  
For very different effective masses and
large enough tunneling, such as $M=10$, $t=5$, two levels
 can be in the resonant region, such as $E^h_{n+1}$ and $E^h_n$, and two 
anomalous jumps in each layer can be observed. 
Naturally, the anomalous part of a jump
(different from 1 or 0 conductivity quantum) decreases away from the
resonant conditions. Experimentally, the tuning of the resonance
 can be achieved in principle by changing the bias voltage $V_0$
(or energy offset $E_0$ in our description).  The sign of the jump of
each component at a particular resonance is defined by the relative
position of the resonant LL's.  For example, in the case $E_n^l\approx
E_{n+1}^h$, the jump of the component $\sigma_{xy}^{ll,ll}$ is negative at
$E_{n+1}^h<E_n^l$, and it is positive for the opposite relation. Weak
disorder in each layer softens the shape of the jumps.

For strong disorder, the situation changes even more, as it influences
the position of the energy levels and the density of states.  As mentioned
above, the disorder results in broadening, given mathematically by the
appearance of an imaginary part for each energy, e.g., $E_k^l\rightarrow
E_k^l- i\frac{q}{2\tau_l}$, which results in a shift and
the appearance of imaginary parts in the energies $\epsilon^\pm$ in
Eq.\ (\ref{shiftLL}).  For strong disorder,
 the energies $\epsilon^{\pm}$ exhibit square root singularities as functions 
of the Fermi energy. These 
in turn result in singular behavior of the density of states, and
in the appearance of additional features in
$\sigma_{xy}^{ll,ll}$, $\sigma_{xy}^{hh,hh}$, and other components in the 
case {\bf (i)} (not shown).  As 
for other jumps, the combination in  (\ref{quant}) remains continuous 
and with well-quantized values, 
while the individual components have unusual
features.  For the resonance between the LL's with different quantum
numbers, case {\bf (ii)}, the strong disorder produces just increasing
broadening of each level.  

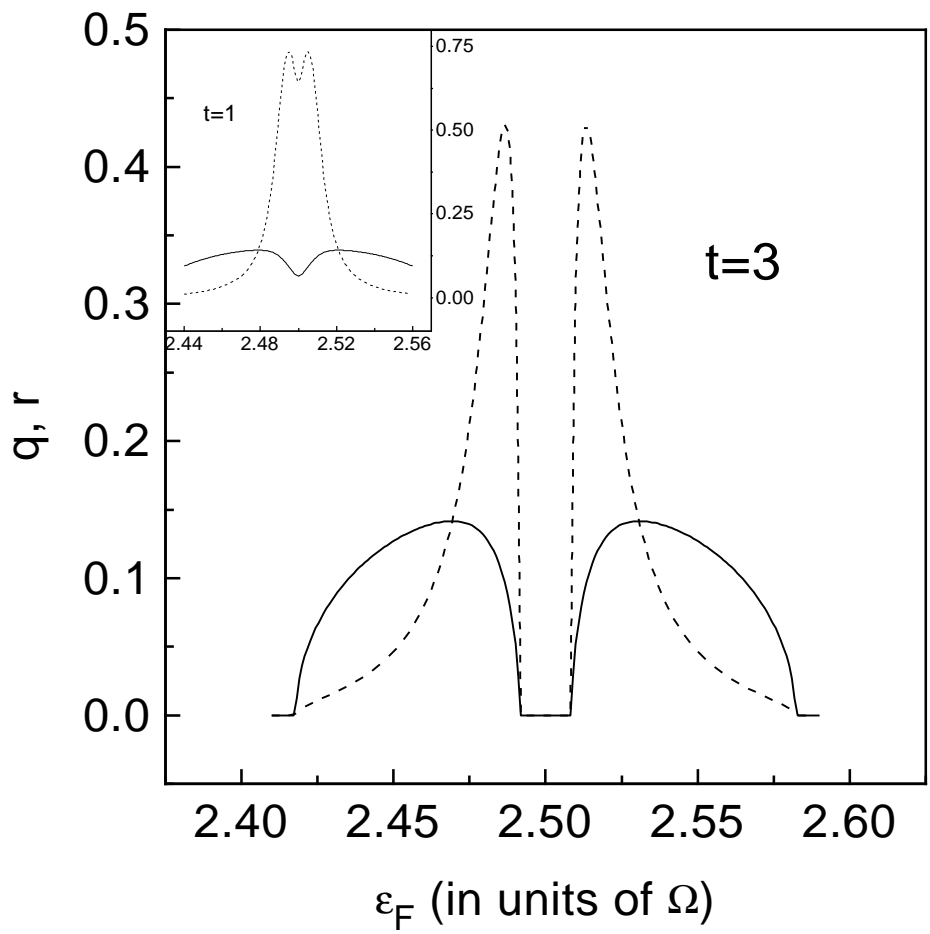
\begin{figure}
\caption{Typical solutions to the mean field equations, 
$q(\epsilon_F)$ (solid line) and $r(\epsilon_F)$ (dashed).  Here,
 $\tau_l=0.01$, $\tau_h=1$, $M=5$, and $t=3$.
The inset shows a case of smaller tunneling, $t=1$.
\label{figres2}
}
\end{figure}

We should notice that the solutions of the nonlinear equations 
for $q$ and $r$ (see Eq.\ (\ref{q})) are related to the structure 
of the density of states
of the coupled system.  The values of $q$ and $r$ change with the
position of the Fermi energy, as shown in Fig.\ \ref{figres2}, for a
typical case.  Note that the gaps in $q(\epsilon_F)$ and
$r(\epsilon_F)$ have the same origin as the gap in the density of
states found for a 2DEG--QD system in Ref. \cite{Shahbazyan}.  Our
results here should then provide a description of the transport properties
for such a system \cite{Ensslin}.

It is important to notice that all of these results for
$\hat{\sigma}_{xy}$, which have been obtained by treating the disorder
on a mean-field level, are not affected by the fluctuations.  To study
this, we have also derived the nonlinear $\sigma$-model for the
two-layer QH system. Details will be presented elsewhere \cite{Alextodo}.
The fluctuation expansion around the mean-field
solution leads to a nonlinear $\sigma$-model of the same class that the
model for the single-layer system \cite{Pruisken}.  Here, the role of
the coupling constants is played by the traces of longitudinal and
transverse conductivity tensors.  Analysis of the Ward identities shows
that the diagonal components of the transverse conductivity tensor are
coupled to a topologically invariant term only \cite{Pruisken}, and are
therefore not affected by the fluctuations.  As such, one finds the identity
of the loop expansions for each component of the tensor
$\hat{\sigma}_{xy}$, and for the transverse
conductivity of the one layer system, up to a general prefactor.  This in turn
 leads one to the conclusion that
the results for each component of the transverse conductivity tensor,
obtained at the mean-field level, are robust against the fluctuations.

We thank P. Pereyra for fruitful discussions. This work was supported 
in part by US Department of Energy Grant No.\ DE-FG02-91ER45334.

\begin{figure}
\caption{Experimental scheme for the measurements of the quantum Hall 
conductivity resonances.  The voltage $V_0$ 
shifts LL ladders in the $l$- and $h$-layers relative to one another.}
\label{figexp}
\end{figure}
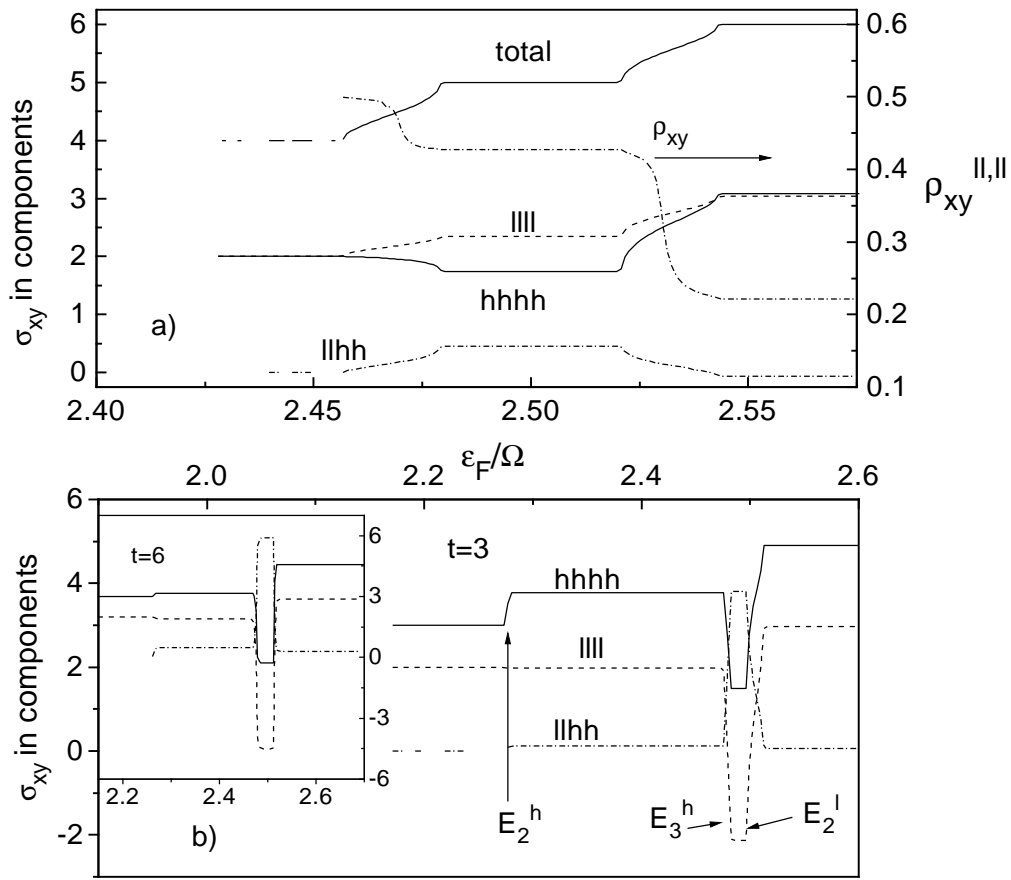
\begin{figure}
\caption{The transverse conductivity components vs Fermi energy at weak disorder, 
$\tau_l=1$, $\tau_h=10$,  $M=5$, $t=3$, $\Omega=100$. The total conductivity 
corresponds to the expression (\ref{quant}). 
(a) The case of the resonance  $E_2^h=E_2^l=5\Omega/2$.
The dashed-dotted line shows the component $\rho_{xy}^{ll,ll}$ of the resistivity 
matrix, which illustrates the nonquantized plateaus of the resistivity. 
(b) The case of the resonance  $E_3^h+\delta=E_2^l$.  Arrows show the 
correspondence 
between the positions of the jumps and the energies of unshifted LL's. 
In the inset, a large tunneling case, $t=6$, is shown. 
The resonant region is larger and involves also the level $E_2^h$.
\label{figres1}
}
\end{figure}

\end{document}